\documentclass[aps,showpacs,prl,twocolumn,footinbib]{revtex4-1}
\usepackage{graphicx}% Include figure files
\usepackage{dcolumn}% Align table columns on decimal point
\usepackage{bm}% bold math
\usepackage[colorlinks]{hyperref}
\usepackage[utf8x]{inputenc}
\usepackage{braket}
\usepackage{amssymb}
\usepackage{amsmath}
\usepackage{bibentry}
\usepackage{graphicx}
\usepackage[colorinlistoftodos]{todonotes}
\usepackage{lipsum}

\begin{document}

\preprint{APS/123-QED}

\title{Dynamical creation and detection of entangled many-body states in a chiral atom chain}% Force line breaks with \\

\author{Giuseppe Buonaiuto}
\email{giuseppe.buonaiuto@nottingham.ac.uk}
\affiliation{School of Physics and Astronomy and Centre for the Mathematics and Theoretical Physics of Quantum Non-Equilibrium Systems, University of Nottingham, Nottingham NG7 2RD, United Kingdom}
\author{Ryan Jones}
\affiliation{School of Physics and Astronomy and Centre for the Mathematics and Theoretical Physics of Quantum Non-Equilibrium Systems, University of Nottingham, Nottingham NG7 2RD, United Kingdom}
\author{Beatriz Olmos}
\affiliation{School of Physics and Astronomy and Centre for the Mathematics and Theoretical Physics of Quantum Non-Equilibrium Systems, University of Nottingham, Nottingham NG7 2RD, United Kingdom}
\author{Igor Lesanovsky}
\affiliation{School of Physics and Astronomy and Centre for the Mathematics and Theoretical Physics of Quantum Non-Equilibrium Systems, University of Nottingham, Nottingham NG7 2RD, United Kingdom}

\begin{abstract}
Open quantum systems with chiral interactions can be realized by coupling atoms to guided radiation modes in waveguides or optical fibres. In their steady state these systems can feature intricate many-body phases such as entangled dark states, but their detection and characterization remains a challenge. Here we show how such collective phenomena can be uncovered through monitoring the record of photons emitted into the guided modes. This permits the identification of dark entangled states but furthermore offers novel capabilities for probing complex dynamical behavior, such as the coexistence of a dark entangled and a mixed phase. Our results are of direct relevance for current experiments, as they provide a framework for probing, characterizing and classifying dynamical features of chiral light-matter systems.
\end{abstract}
\date{\today}

\maketitle
\textit{Introduction}.
The ability to interface quantum emitters with optical systems opens novel routes for investigating non-equilibrium phenomena in open condensed matter physics \cite{Lee} and provides, potentially, a platform to perform quantum information processing \cite{Weim,Stann,Lvov,Asen}. In recent years, the open quantum dynamics of chiral systems, where the emission of photons into a waveguide presents a broken left-right symmetry, has been the object of intense investigation \cite{Dutta, Scheu, Lod, Verm,Arno2}. This propagation-direction-dependent light-matter interaction has been observed in a variety of systems, for instance atoms coupled to the evanescent field of a waveguide \cite{Say}, and quantum dots in photonic nano-structures \cite{Coles}.

From the quantum many-body physics perspective, it has been demonstrated that chiral coupling induces steady state entanglement between emitters. In particular, when a chain of spins is coupled to a waveguide, under precise conditions, peculiar phases of matter can form, among them pure entangled many-body dark states that are protected from decoherence \cite{Pich}. However, it still remains an experimental challenge to realize the specific conditions required for the creation of these entangled dark states. Furthermore, it is generally a difficult task to detect and characterize such states once created. This poses obstacles for exploring and using of the rich variety of quantum states and many-body phases for practical purposes, e.g. in optical quantum computing protocols \cite{Met}.

In this work, we show that the counting statistics of the photons emitted into the waveguide provides a means to detect and characterize the many-body state of a chiral atom chain. Our approach exploits that the guided photons do not only induce interactions between the emitters, but also carry information about their quantum state. In an experiment, measuring the output field of the waveguide thus allows to infer the occurrence of dark steady states. More generally, this perspective allows to probe intricate dynamical phenomena such as transitions and the coexistence between dynamical phases in the steady state. In the latter case, entangled states may occur as fluctuations in an intermittent dynamics and are heralded through characteristic features of the time-resolved photon count signal. To establish this correspondence on a formal level we utilize the scaled cumulant generating function formalism together with the large deviation principle \cite{touchette2009,Garr}. This does not only allow to develop an understanding of the dynamical non-equilibrium behavior of a chirally coupled atom chain, but also permits to systematically assess the effect of inevitable imperfections, such as the emission of photons into unguided modes.

\begin{figure}
\centering
\includegraphics[scale=0.29]{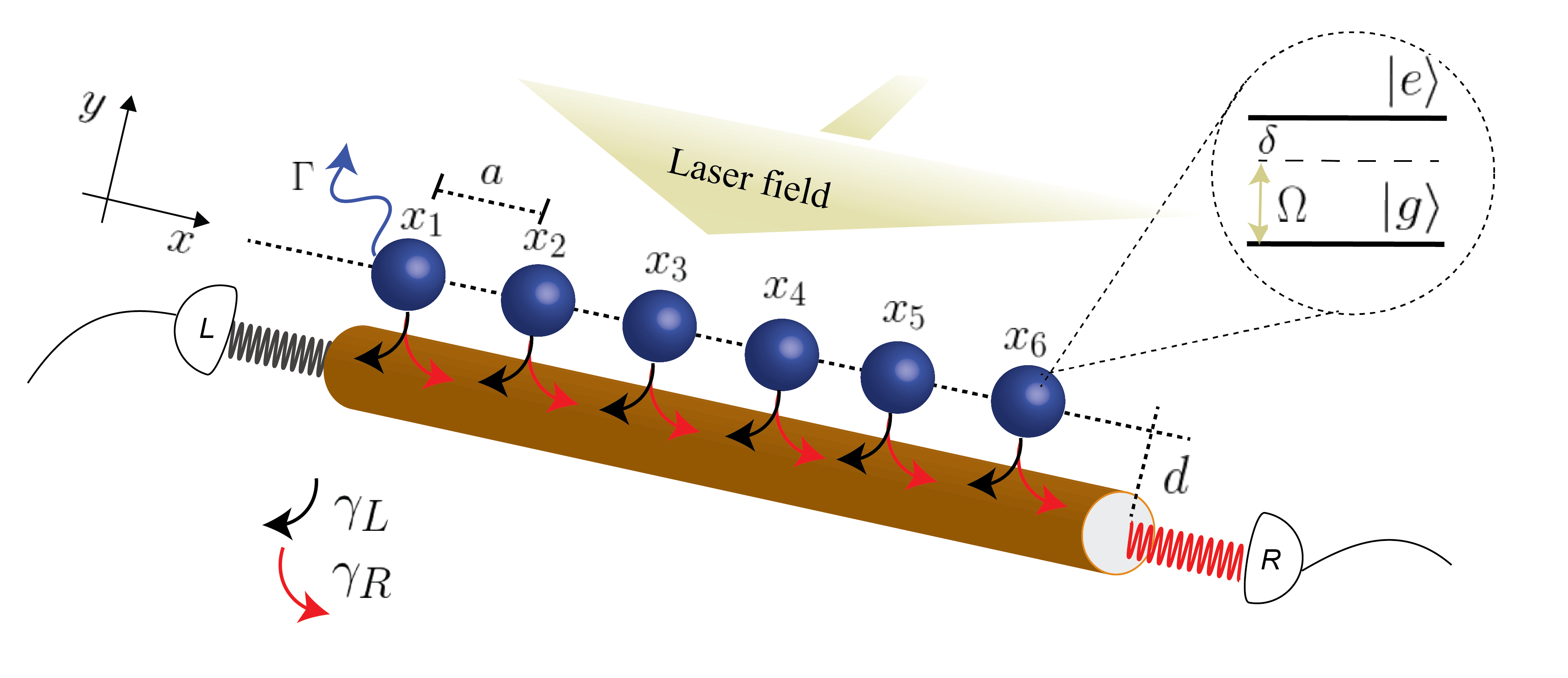}
\caption{\textbf{Atom chain with chiral interactions.} A chain of two-level atoms, separated from each other by a distance $a$, is placed at a distance $d$ from the center of a waveguide. The system is driven homogeneously via a laser field, with Rabi frequency $\Omega$ and detuning $\delta$. The photons emitted into the left and right guided modes of the waveguide with rates $\gamma_L$ and $\gamma_R$, respectively, are collected by two detectors placed at the ends of the waveguide. Emission of unguided photons into unguided modes occurs at a rate $\Gamma$.}\label{scheme}
\end{figure}

\textit{Model}. The general setup that we consider is depicted in Fig. \ref{scheme}. Here, $N$ spin-$1/2$ particles form a chain parallel to a one-dimensional waveguide, placed at positions $x_{j}$, with $j=1,\dots,N$. Note that in the following we explicitly refer to two-level atoms with internal states $\ket{g}$ and $\ket{e}$, but the same description applies, e.g. to quantum dots in photonic crystals. The atoms are trapped at a distance $d$ from the center of the waveguide and separated by a distance $a$ from each other. Each atom is driven by a coherent driving field with frequency $\omega_{p}$, Rabi frequency $\Omega$, and a (possibly spatially dependent) detuning $\delta_{j}=\omega_{p}-\omega_{j}$, with $\omega_{j}$ being the transition frequency between the ground and the excited state. The Hamiltonian that governs the dynamics of the laser excitation (under the rotating wave approximation) is $H_\textrm{laser}=\sum_{j=1}^{N}\left[-\delta_{j}\sigma^{+}_{j}\sigma^{-}_{j}+\Omega(\sigma^{+}_{j}+\sigma^{-}_{j})\right]$, where $\sigma^{-}_{j}=\ket{g}_{j}\!\bra{e}$.

The atoms are coupled both to the left and right propagating \textit{guided} modes of the waveguide as well as to the \textit{unguided} modes of the electromagnetic field outside the waveguide. Two photon detectors are placed at the extremes of the waveguide which allows in principle to resolve the photons emitted into both directions. Under the Born-Markov approximation, the open quantum dynamics of the atoms can be described by a Lindblad master equation:
\begin{equation}
\label{mastergen}
\begin{split}
\dot{\rho}=&\mathcal{W}(\rho)=-i[H_\textrm{laser}+H_{L}+H_{R},\rho]\\
& +\gamma_{L}\mathcal{D}(J_{L})\rho+\gamma_{R}\mathcal{D}(J_{R})\rho+\Gamma\sum_{j=1}^N\mathcal{D}(\sigma_j^-)\rho.
\end{split}
\end{equation}
Here, $H_L$ and $H_R$ describe induced coherent dipole-dipole interactions due to the coupling between the atom chain and the left and right propagating modes of the waveguide, respectively. They read
\begin{equation}
\begin{split}
&H_{L}=-i\frac{\gamma_{L}}{2}\sum_{j<l}(e^{-ik|x_{j}-x_{l}|}\sigma^{+}_{j}\sigma^{-}_{l}+\textrm{h.c}) \\ & H_{R}=-i\frac{\gamma_{R}}{2}\sum_{j>l}(e^{-ik|x_{j}-x_{l}|}\sigma^{+}_{j}\sigma^{-}_{l}+\textrm{h.c}),
\end{split}
\end{equation}
with $k=\omega_{p}/c$. The incoherent collective spin decay into the guided modes is described by the dissipators $\gamma_{L,R}\mathcal{D}(J_{L,R})$, where $J_{L,R}=\sum_{j}e^{\pm ikx_{j}}\sigma^{-}_{j}$ are jump operators describing the emission of a photon into the guided modes, and $\mathcal{D}(\chi)\rho\equiv\chi\rho\chi^\dagger-\frac{1}{2}\left\{\chi^\dagger\chi,\rho\right\}$. The last term of Eq. \eqref{mastergen} represents the coupling of the atoms to the modes of the free (unguided) electromagnetic field with rate $\Gamma$ \footnote{Note, that this expression is only valid when the separation between the atoms is sufficiently large, $ka\gg1$. When $ka<1$, the emission into the unguided modes becomes collective instead of local, and the exchange of virtual photons gives rise to coherent interactions between the atoms \cite{Lehm}.}.

In the remainder of the paper we assume, for concreteness, that the spacing between the atoms, $a$, is commensurate with the wavelength of the laser field: $\lambda=\frac{2\pi}{k}$, such that $kx_j=2\pi n$ with $n=1,2,\dots$ for all $j=1,\dots,N$. Under these conditions, the master equation \eqref{mastergen} simplifies considerably and takes the form
\begin{equation}
\begin{split}
\label{chiral}
\dot{\rho}=-i\left[H,\rho\right] +\gamma\mathcal{D}(J)\rho+\Gamma\sum_{j=1}^N\mathcal{D}(\sigma_j^-)\rho,
\end{split}
\end{equation}
with $\gamma=\gamma_{R}+\gamma_{L}$ and
\begin{equation*}
H=\Omega\left(J+J^{\dagger}\right)-\sum_{j}\delta_{j}\sigma^{+}_{j}\sigma^{-}_{j}-\frac{i}{2}\Delta\gamma \sum_{j>l}(\sigma^{+}_{j}\sigma^{-}_{l}-\sigma^{+}_{l}\sigma^{-}_{j}),
\end{equation*}
where $J=\sum_{j}\sigma_{j}^{-}$ and $\Delta\gamma=\gamma_{R}-\gamma_{L}$. For now (we will return to this at a later stage) we also neglect decay into the unguided modes, i.e., we consider $\Gamma=0$.

We are here particularly interested in the case of \textit{chiral} interactions, i.e. when the couplings with the left and the right propagating modes are not symmetric and hence $\Delta\gamma\neq0$. This is realized by an appropriate choice of the laser polarization and transition dipole moment of each atom: as shown in Ref. \cite{LeKi}, the dipole moments are required to have a real and imaginary part (elliptically polarized light) in order for the coupling to the guided modes to acquire a chiral character. Note that, in the non-chiral case, i.e. when $\Delta\gamma=0$, Eq. \eqref{chiral}  with $\Gamma=0$ describes a fully symmetric Dicke model \cite{Rod}.

\begin{figure*}[!ht]
\centering
\includegraphics[scale=0.37]{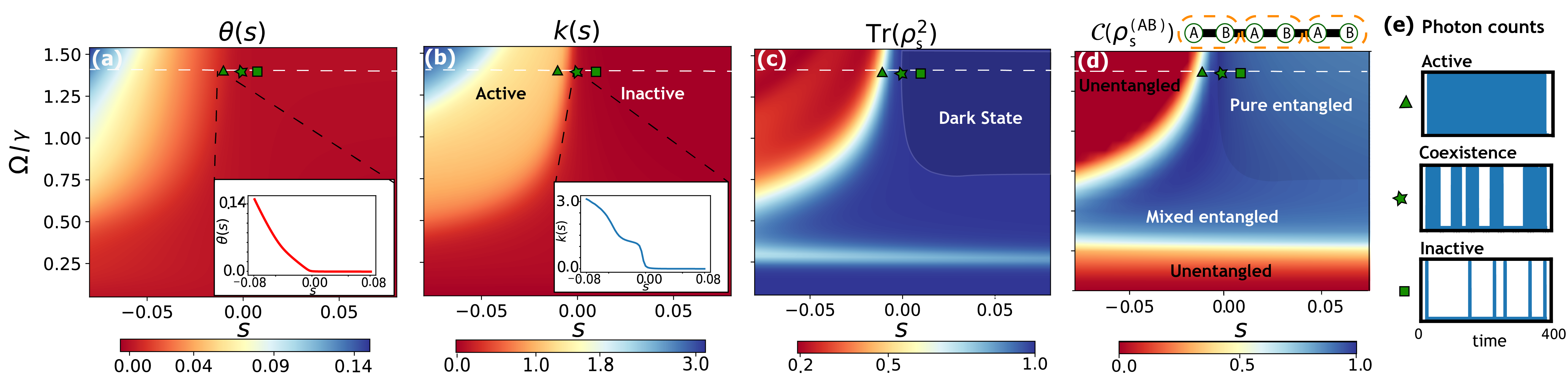}
\caption{\label{nozol} \textbf{Intermittent entangled dark states.} Chain of $N=6$ atoms, with $\Delta\gamma=2\gamma/3$ and spatially uniform detuning $\delta_{j}=\delta=\gamma/10$. (a) Scaled cumulant generating function (SCGF), $\theta(s)$, as function of the Rabi frequency $\Omega$. In the inset, a cut for $\Omega/\gamma=1.4$ (white dashed line in the main panel) shows that the SCGF undergoes a sharp change at $s\approx 0$. (b) Associated photon count $k(s)$, which shows a steep jump at $s=0$ for sufficiently large $\Omega$, more noticeable in the inset for $\Omega/\gamma=1.4$. (c) Purity of the conditioned state $\rho_s$. The state is generally mixed except in a region at large $\Omega$ and $s>0$ (delimited by a fine white line). (d) Concurrence of the reduced two-body conditioned density matrix $\rho_s^{(AB)}$  for the atom pairs shown above the panel. (e) (Conditioned/biased) time-resolved photon emission record calculated at $s=-0.01$, $s=0$ and $s=0.01$ for $\Omega/\gamma=1.4$ (time is in units of $1/\gamma$).}
\end{figure*}

\textit{Photon detection and cumulant generating function}. The statistics of the photon count at the two detectors (see Fig. \ref{scheme}) carries information about the state of the atom chain as well as its dynamical behavior. The counting statistics is fully characterized by the cumulants (e.g. the mean value, the fluctuations, etc.), which are encoded in the scaled cumulant generating function (SCGF). We now briefly summarize how to obtain the SCGF in practice. We start with the density matrix $\rho^{(K_{L},K_{R})}(t)$, which represents the state of the atomic system at time $t$ conditioned to the detection of exactly $K_{L}$ and $K_{R}$ photons on the left and right detector, respectively. The corresponding photon detection probability is thus given by $P_{t}(K_{L},K_{R})=\textrm{Tr}\left[\rho^{(K_{L},K_{R})}(t)\right]$ and its generating function is calculated via the two-dimensional Laplace transform $Z_{t}(s_{L},s_{R})=\sum_{K_{L},K_{R}=0}^{\infty}P_{t}(K_{L},K_{R})e^{-K_{L}s_{L}}e^{-K_{R}s_{R}}$. In the limit of long times, i.e. times much longer than any microscopic time scale of the dynamics, it assumes a so-called large deviation form: $Z_{t}(s_{L},s_{R}) \simeq e^{t\theta(s_{L},s_{R})}$. The SCGF is here given by $\theta(s_{L},s_{R})$, which also happens to be the largest real eigenvalue of the transformed master operator \cite{Garr}
\begin{equation}
\label{Ws}
\mathcal{W}_{s_{L},s_{R}}\!(\bullet)\!=\!\mathcal{W}(\bullet)
-J_{L}\bullet J^{\dagger}_{L}(1\!-e^{-s_{L}})-J_{R}\bullet J^{\dagger}_{R}(1\!-e^{-s_{R}}).
\end{equation}
This permits a direct calculation of the SCGF, whose derivatives with respect to $s_L$ and $s_R$, taken at $s_L=s_R=0$, yield the moments of the photon number distribution. For example, the average number of photons emitted into the left propagating mode is given by $k_{L}=-\partial_{s_L}\theta(s_{L},s_{R})\!\mid_{s_L=s_R=0}$, while the corresponding fluctuations are given by $\partial^{2}_{s_{L}}\theta(s_{L},s_{R})\!\mid_{s_L=s_R=0}$. Hence, a sudden change of the structure of the SCGF in the vicinity of $s_L=s_R=0$ is concomitant with large cumulants, e.g. strong fluctuations, which is a manifestation of non-trivial dynamical behavior \cite{Bark,IgMad}.

\textit{Intermittent entangled dark states}. To be more concrete we consider a system of $N=6$ atoms in the chiral regime ($\Delta\gamma \neq 0$), described by the master equation \eqref{chiral} with $\Gamma=0$. All atoms are driven by a laser field with spatially uniform detunings $\delta_{j}=\delta$. To simplify things further, we count the total emission $K\!=\!K_L\!+\!K_R$ into both detectors, without resolving the left and right photons separately  [$s_{L}\!=\!s_{R}\!=\!s$, $\theta(s_{L},s_{R})\rightarrow\theta(s)$]. The resulting SCGF is displayed in Fig. \ref{nozol}(a) as a function of the laser Rabi frequency $\Omega$. For sufficiently large $\Omega$, $\theta(s)$ shows a sudden change around $s=0$. This signals a sharp increase of the fluctuations in the number of photons arriving at the detectors per unit of time.

We show below that this is a consequence of the fact that the stationary state for large $\Omega$ is actually one where two states (or phases) with drastically different dynamical properties coexist. To analyze the nature of this phenomenon, let us come back to the generating function of the photon count distribution function:
\begin{equation}
\label{field}
Z_{t}(s)\!=\!\sum_{K}\textrm{Tr}\!\left[\rho^{(K)}\!\right]\!e^{-sK}\!=\!\textrm{Tr}\!\left[\!\sum_{K}\rho^{(K)}\!e^{-sK}\!\right]\!\!=\!\textrm{Tr}\!\left[\rho_s\right].
\end{equation}
According to this formula, $Z_{t}(s)$ is given as the trace of a biased (unnormalized) density matrix $\rho_s$: For $s<0$, it is biased towards conditional states, $\rho^{(K)}$, with large numbers of photon count $K$, while for $s>0$ the states with low $K$ are the ones dominantly contributing to $\rho_s$. Note that $\rho_s$ can in practice be found as the eigenstate of the dynamical generator given by Eq. \eqref{Ws} corresponding to the largest eigenvalue $\theta(s)$.

The fact that the SCGF changes quickly in the vicinity of $s=0$, as seen in Fig. \ref{nozol}(a), therefore signals that the biased state $\rho_s$ changes drastically when $s$ is varied across zero. In the case at hand the steady state turns out to be characterized by the coexistence of a dark state with zero photon count $K$ (inactive phase) and a bright (active) phase with large $K$. This is seen in Fig. \ref{nozol}(b), where we show the (biased) mean number of emitted photons per unit time, $k(s)$ \footnote{Note here that, for $s<0$, we find another crossover in $k(s)$: This is a finite-size effect as we confirmed by numerical analysis.}.

Analyzing the biased state $\rho_s$ for positive and negative values of $s$ provides direct access to the properties of the two phases that coexist in the steady state. The inactive phase ($\rho_s$ for $s>0$) is characterized by the absence of photon emission into the waveguide. It is expected to be a so-called \textit{dark state} \cite{Hu}, which is pure and in which "photons are trapped" in the atomic system due to quantum interference such that the steady state is virtually uncoupled from the modes of the waveguide. Indeed, calculating the purity of the biased state, $\mathrm{Tr}\left(\rho^{2}_s\right)$ [see Fig. \ref{nozol}(c)], shows that for large $\Omega$ and $s>0$ a pure state is assumed. Conversely, a mixed state is found for $s<0$.
Interestingly, the dark state is entangled in a very peculiar way: In Fig. \ref{nozol}(d) we show the concurrence \cite{Woot} ${\cal C}\left(\rho_s^{(AB)}\right)$ of the two-atom reduced (and biased) density matrix $\rho_s^{(AB)}$ for the $AB$-pairs shown above the panel. Note, that other pairs are not entangled, i.e. this inactive phase is one of highly entangled dimers.

Fig. \ref{nozol} nicely illustrates the strength of the applied analysis method: large fluctuations, i.e. a steep change of the photon count rate $k(s)$ at $s=0$, are indeed the consequence of two coexisting phases in the steady state. Experimentally, this manifests itself in a highly intermittent photon count, as shown in Fig. \ref{nozol}(e) (middle panel). However, the theoretical investigation of the biased states $\rho_s$ reveals much more. It shows that with increasing $\Omega$ the steady state of the chiral atom chain is at first unimodal and enters a mixed entangled phase from an unentangled pure (product) state. Increasing $\Omega$ further, the steady state acquires an at first highly suppressed (only appearing at large negative $s$) component of a mixed phase. The weight of this phase increases until it coexists with the entangled phase, which has in turn become a pure dark state.

\begin{figure}
\includegraphics[scale=0.31]{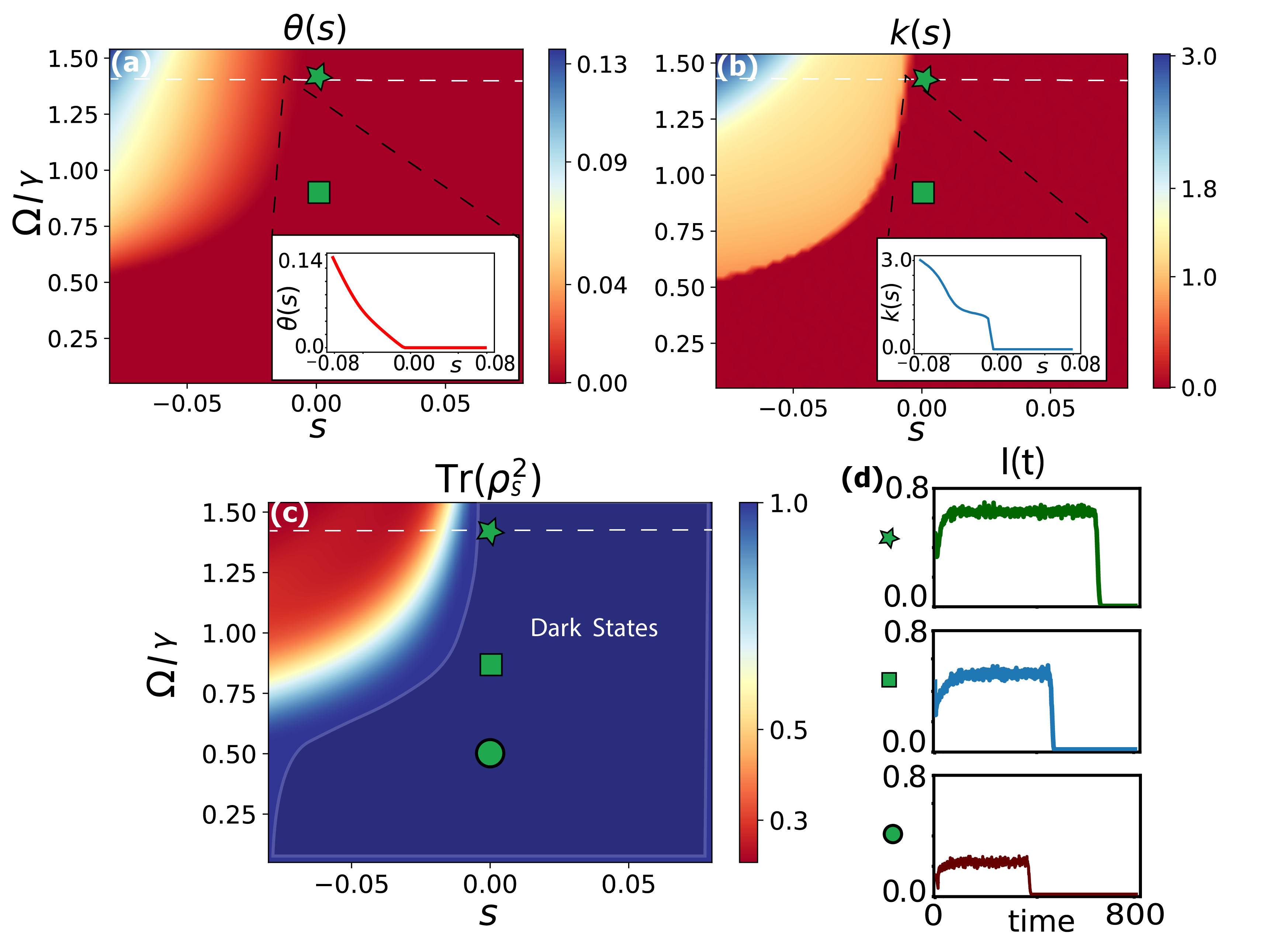}
\caption{\label{zol}\textbf{Steady state dark states.} The system parameters are the same as in Fig. \ref{nozol}, but the detuning is alternating, $\delta_{j}=(-1)^{j}\gamma/10$. (a) SCGF, $\theta(s)$, as a function of the Rabi frequency $\Omega$. In the inset, a cut for $\Omega/\gamma=1.4$ (white dashed line in the main panel) shows that the SCGF has a sharp change this time at $s<0$, i.e. active and inactive phase do not coexist. (b) (Biased) steady state photon count rate $k(s)$ and cut taken at $\Omega/\gamma=1.4$. (c) Purity of the biased state $\rho_s$. The dark entangled state $\rho_\mathrm{dimer}$ is the stationary state of the system for all $\Omega$. (d) The photo current $I(t)$ shows that the dark state is indeed reached after a ($\Omega$-dependent) transient period. The initial state chosen here is one where all atoms are in their ground state $\left|g\right>$ and the time is in units of $1/\gamma$.}
\end{figure}

We now turn briefly to a case which was recently investigated in Ref. \cite{Pich}. Here the atoms are excited with a laser field whose detuning is alternating, $\delta_{j}=(-1)^j\delta$. This may be challenging to achieve experimentally but has the advantage that the steady state can be calculated analytically: it is a dark state formed by dimers, which takes the form $\rho_\mathrm{dimer}=\ket{\psi}\bra{\psi}$, where $\ket{\psi}=\bigotimes_{j=1}^{N/2}\ket{D}_{2j-1,2j}$ and $\ket{D}\propto\ket{gg}+\beta\left(\ket{ge}-\ket{eg}\right)$ with
$\beta=-2\Omega/(2\delta+i\Delta\gamma)$. The corresponding SCGF, the biased photon emission rate as well as the purity of the biased state are shown in Fig. \ref{zol}. The plots are similar to the situation depicted in Fig. \ref{nozol}, i.e. there is an inactive (dark state) phase and an active (mixed) phase that is closing in as $\Omega$ is increased. However, unlike in the previous case, both phases cannot be brought into coexistence, i.e. the mixed phase always remains strongly suppressed. For an initial state in which all atoms are in their ground state $\left|g\right>$ it is only visible in the transient behavior of the photo current, as shown in Fig. \ref{zol}(d).

\begin{figure}
\centering
\includegraphics[scale=0.28]{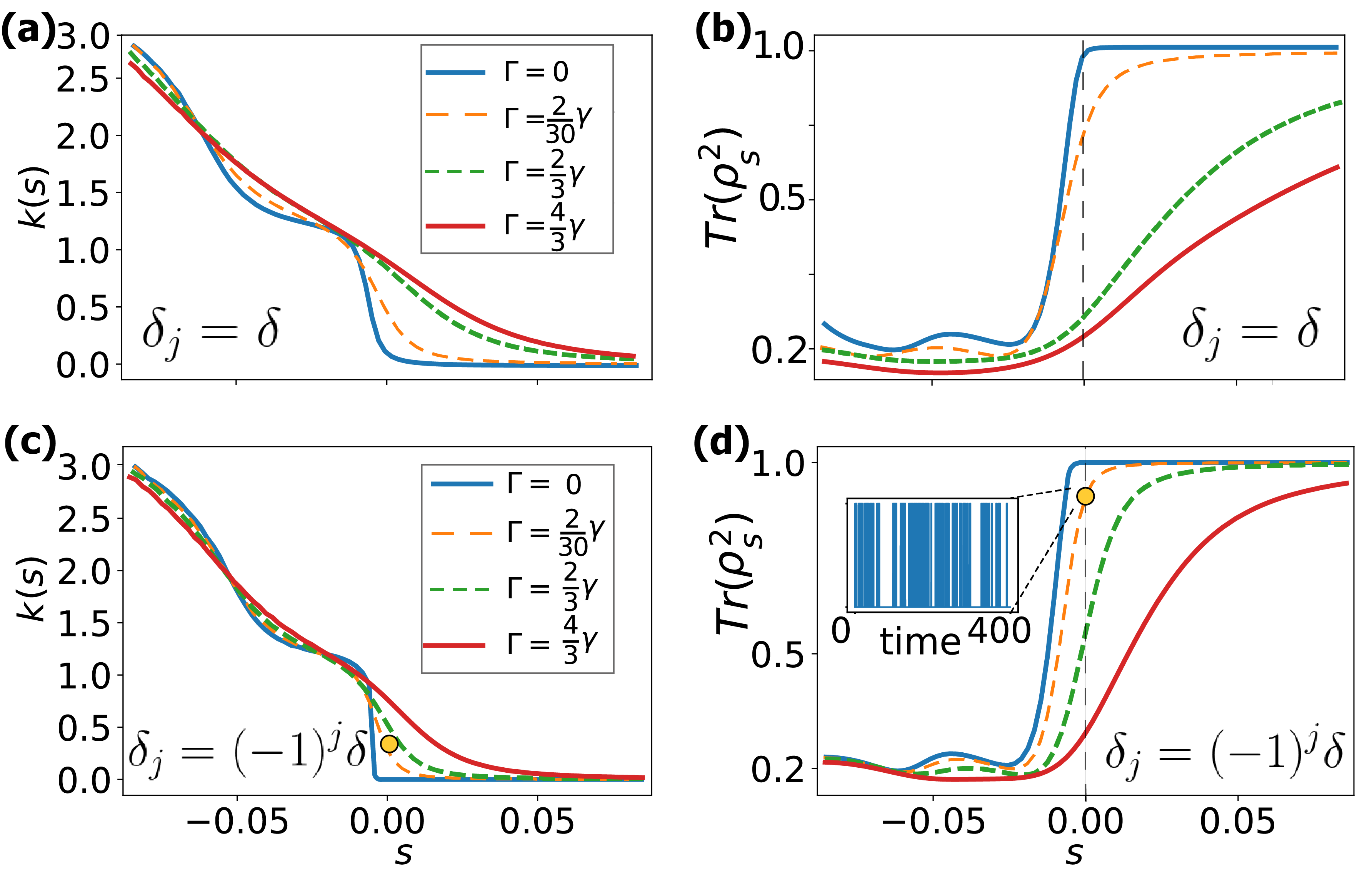}
\caption{\label{unguided} \textbf{Emission into unguided modes.} (a) (Biased) Photon emission rate $k(s)$ for the same system as discussed in Fig. \ref{nozol} [$\delta_j=\delta$] at $\Omega/\gamma=1.4$. Increasing the emission rate $\Gamma$ into unguided modes smoothes out the sharp features of the SCGF. (b) Purity of the biased state $\rho_s$. Increasing $\Gamma$ moves the steady state into a mixed phase. (c) (Biased) Photon emission rate $k(s)$ for the same system as discussed in Fig. \ref{zol} [$\delta_j=(-1)^j\delta$] at $\Omega/\gamma=1.4$. (d) Purity of the biased state $\rho_s$. Inset: Unbiased photon emission record (for $\Gamma=2\gamma/30$) showing intermittent behaviour due to phase coexistence.}
\end{figure}

\textit{Emission into unguided modes}. In an experiment, photons are inevitably emitted also into unguided modes \cite{Gob} at a finite rate $\Gamma$. We investigate this effect in Fig. \ref{unguided} for the two detuning patterns discussed in Fig. \ref{nozol} [$\delta_j=\delta$] and Fig. \ref{zol} [$\delta_j=(-1)^j\delta$], respectively. We find that an increasing value of $\Gamma$ generally smoothes the sharp features in the (biased) emission rate $k(s)$ and that the mixedness of the steady state increases. While the behavior of the case with homogeneous detuning [Figs. \ref{unguided}(a,b)] is only qualitatively affected, the situation changes quite drastically in the alternating case [Figs. \ref{unguided}(c,d)]. Here we see that a small $\Gamma$ removes the exact dark steady state and brings it into coexistence with a mixed phase. The result is an intermittent photon emission dynamics which is no longer transient, as is shown in the inset of Fig. \ref{unguided}(d). I.e. just like in Fig. \ref{nozol} the pure dimer dark state is probed dynamically through fluctuations while the stationary state becomes mixed.

Finally we remark, that the coupling strength between the atoms (or generically the quantum emitters) and the guided modes can be controlled by employing a cavity, which is coupled itself to the guided modes. In this way, as showed in \cite{Tiec}, it is possible to engineer a coupling to the unguided modes of one order of magnitude smaller than the one to the guided ones.

\textit{Conclusion}. The counting statistics of photons emitted into guided modes yields direct insights into the steady state dynamics of chiral atom chains. This connection does not only allow to infer the emergence of dark state phases, but --- more importantly --- provides also in-situ access to dynamical properties and fluctuations. We demonstrated this by focussing on a regime of coexistence between an entangled dark state and a mixed bright phase. Beyond its appeal from an experimental perspective, our approach also shows the conceptual strength of the SCGF method and the usefulness of the correspondingly biased quantum state for unravelling complex dynamical behavior in open quantum systems. In the future it would be interesting to explore other photon detection protocols, e.g. homodyne detection, and the possibility to directly create the biased quantum state via post-selection \cite{Gamm,Tan} or a "deformed dynamics" \cite{Fed}. The latter is particularly relevant when considering the emission into unguided modes, where dark entangled phases merely emerge as a fluctuation in the steady state dynamics.

\begin{acknowledgements}
We thank Dario Cilluffo and all members of the ErBeStA consortium for fruitful discussions. The research leading to these results has received funding from the European Research Council under the European Union's Seventh Framework Programme (FP/2007-2013)  [ERC Grant Agreement No. 335266 (ESCQUMA)] and from the European Union's H2020 research and innovation programme [Grant Agreement No. 800942 (ErBeStA)]. Funding was also received from the EPSRC [Grant No. EP/M014266/1]. IL gratefully acknowledges funding through the Royal Society Wolfson Research Merit Award. BO was supported by the Royal Society and EPSRC [Grant No. DH130145].
\end{acknowledgements}

\bibliographystyle{ieeetr}

\bibliography{papp}

\begin{thebibliography}{10}

\bibitem{Lee}
K.~H. Lee, C.~Lee, H.~Min, and S.~B. Chung, ``Phase transitions of the
  polariton condensate in 2d dirac materials,'' {\em Phys. Rev. Lett.},
  vol.~120, p.~157601, 2018.

\bibitem{Weim}
H.~Weimer, M.~M{\"u}ller, I.~Lesanovsky, P.~Zoller, and H.~P. B{\"u}chler, ``A
  rydberg quantum simulator,'' {\em Nat. Phys.}, vol.~6, p.~382, 2010.

\bibitem{Stann}
K.~Stannigel, P.~Rabl, A.~S. S\o{}rensen, M.~D. Lukin, and P.~Zoller,
  ``Optomechanical transducers for quantum-information processing,'' {\em Phys.
  Rev. A}, vol.~84, p.~042341, 2011.

\bibitem{Lvov}
A.~I. Lvovsky, B.~C. Sanders, and W.~Tittel, ``{Optical quantum memory},'' {\em
  Nat. Phot.}, vol.~3, p.~706, 2009.

\bibitem{Asen}
A.~Asenjo-Garc\'{i}a, M.~Moreno-Cardoner, A.~Albrecht, H.~J. Kimble, and D.~E.
  Chang, ``Exponential improvement in photon storage fidelities using
  subradiance and ``selective radiance'' in atomic arrays,'' {\em Phys. Rev.
  X}, vol.~7, p.~031024, 2017.

\bibitem{Dutta}
F.~Le~Kien, S.~D. Gupta, K.~P. Nayak, and K.~Hakuta, ``Nanofiber-mediated
  radiative transfer between two distant atoms,'' {\em Phys. Rev. A}, vol.~72,
  p.~063815, 2005.

\bibitem{Scheu}
M.~Scheucher, A.~Hilico, E.~Will, J.~Volz, and A.~Rauschenbeutel, ``Quantum
  optical circulator controlled by a single chirally coupled atom,'' {\em
  Science}, vol.~354, no.~6319, p.~1577, 2016.

\bibitem{Lod}
P.~Lodahl, S.~Mahmoodian, S.~Stobbe, A.~Rauschenbeutel, P.~Schneeweiss,
  J.~Volz, H.~Pichler, and P.~Zoller, ``Chiral quantum optics,'' {\em Nature},
  vol.~541, p.~473, 2017.

\bibitem{Verm}
B.~Vermersch, P.-O. Guimond, H.~Pichler, and P.~Zoller, ``Quantum state
  transfer via noisy photonic and phononic waveguides,'' {\em Phys. Rev.
  Lett.}, vol.~118, p.~133601, 2017.

\bibitem{Arno2}
P.~Schneeweiss, S.~Zeiger, T.~Hoinkes, A.~Rauschenbeutel, and J.~Volz, ``Fiber
  ring resonator with a nanofiber section for chiral cavity quantum
  electrodynamics and multimode strong coupling,'' {\em Opt. Lett.}, vol.~42,
  no.~1, p.~85, 2017.

\bibitem{Say}
C.~Sayrin, C.~Junge, R.~Mitsch, B.~Albrecht, D.~O'Shea, P.~Schneeweiss,
  J.~Volz, and A.~Rauschenbeutel, ``Nanophotonic optical isolator controlled by
  the internal state of cold atoms,'' {\em Phys. Rev. X}, vol.~5, p.~041036,
  2015.

\bibitem{Coles}
R.~J. Coles, D.~M. Price, J.~E. Dixon, B.~Royall, E.~Clarke, P.~Kok, M.~S.
  Skolnick, A.~M. Fox, and M.~N. Makhonin, ``Chirality of nanophotonic
  waveguide with embedded quantum emitter for unidirectional spin transfer,''
  {\em Nat. Comm.}, vol.~7, p.~11183, 2016.

\bibitem{Pich}
H.~Pichler, T.~Ramos, A.~J. Daley, and P.~Zoller, ``Quantum optics of chiral
  spin networks,'' {\em Phys. Rev. A}, vol.~91, p.~042116, 2015.

\bibitem{Met}
A.~Metelmann and A.~A. Clerk, ``Nonreciprocal photon transmission and
  amplification via reservoir engineering,'' {\em Phys. Rev. X}, vol.~5,
  p.~021025, 2015.

\bibitem{touchette2009}
H.~Touchette, ``The large deviation approach to statistical mechanics,'' {\em
  Phys. Rep.}, vol.~478, no.~1-3, pp.~1--69, 2009.

\bibitem{Garr}
J.~P. Garrahan and I.~Lesanovsky, ``Thermodynamics of quantum jump
  trajectories,'' {\em Phys. Rev. Lett.}, vol.~104, p.~160601, 2010.

\bibitem{Note1}
Note, that this expression is only valid when the separation between the atoms
  is sufficiently large, $ka\gg 1$. When $ka<1$, the emission into the unguided
  modes becomes collective instead of local, and the exchange of virtual
  photons gives rise to coherent interactions between the atoms \cite {Lehm}.

\bibitem{LeKi}
F.~Le~Kien and A.~Rauschenbeutel, ``Nanofiber-mediated chiral radiative
  coupling between two atoms,'' {\em Phys. Rev. A}, vol.~95, p.~023838, 2017.

\bibitem{Rod}
J.~P.~J. Rodriguez, S.~A. Chilingaryan, and B.~M. Rodr\'{\i}guez-Lara,
  ``Critical phenomena in an extended dicke model,'' {\em Phys. Rev. A},
  vol.~98, p.~043805, 2018.

\bibitem{Bark}
E.~Barkai, Y.~Jung, and R.~J. Silbey, ``Theory of single-molecule spectroscopy:
  beyond the ensemble average.,'' {\em Annu. Rev. Phys. Chem.}, vol.~55,
  p.~457, 2004.

\bibitem{IgMad}
I.~Lesanovsky, M.~van Horssen, M.~Gu\ifmmode \mbox{\c{t}}\else
  \c{t}\fi{}\ifmmode~\u{a}\else \u{a}\fi{}, and J.~P. Garrahan,
  ``Characterization of dynamical phase transitions in quantum jump
  trajectories beyond the properties of the stationary state,'' {\em Phys. Rev.
  Lett.}, vol.~110, p.~150401, 2013.

\bibitem{Note2}
Note here that, for $s<0$, we find another crossover in $k(s)$: This is a
  finite-size effect as we confirmed by numerical analysis.

\bibitem{Hu}
Z.~Hu, G.~S. Engel, F.~H. Alharbi, and S.~Kais, ``Dark states and
  delocalization: Competing effects of quantum coherence on the efficiency of
  light harvesting systems,'' {\em J. Chem. Phys.}, vol.~148, no.~6, p.~064304,
  2018.

\bibitem{Woot}
W.~K. Wootters, ``Entanglement of formation of an arbitrary state of two
  qubits,'' {\em Phys. Rev. Lett.}, vol.~80, p.~2245, 1998.

\bibitem{Gob}
A.~Goban, C.-L. Hung, J.~D. Hood, S.-P. Yu, J.~A. Muniz, O.~Painter, and H.~J.
  Kimble, ``Superradiance for atoms trapped along a photonic crystal
  waveguide,'' {\em Phys. Rev. Lett.}, vol.~115, p.~063601, 2015.

\bibitem{Tiec}
T.~G. Tiecke, J.~D. Thompson, N.~P. de~Leon, L.~R. Liu, V.~Vuletic, and M.~D.
  Lukin, ``Nanophotonic quantum phase switch with a single atom,'' {\em
  Nature}, vol.~508, p.~241, 2014.

\bibitem{Gamm}
S.~Gammelmark, B.~Julsgaard, and K.~M\o{}lmer, ``Past quantum states of a
  monitored system,'' {\em Phys. Rev. Lett.}, vol.~111, p.~160401, 2013.

\bibitem{Tan}
D.~Tan, N.~Foroozani, M.~Naghiloo, A.~H. Kiilerich, K.~M\o{}lmer, and K.~W.
  Murch, ``Homodyne monitoring of postselected decay,'' {\em Phys. Rev. A},
  vol.~96, p.~022104, 2017.

\bibitem{Fed}
F.~Carollo, J.~P. Garrahan, I.~Lesanovsky, and C.~P\'erez-Espigares, ``Making
  rare events typical in markovian open quantum systems,'' {\em Phys. Rev. A},
  vol.~98, p.~010103, 2018.

\bibitem{Lehm}
R.~H. Lehmberg, ``Radiation from an $n$-atom system. i. general formalism,''
  {\em Phys. Rev. A}, vol.~2, p.~883, 1970.

\end{thebibliography}

\end{document}